\documentclass[journal]{IEEEtran}
\usepackage{amssymb}
\usepackage{amsbsy}
\usepackage{graphicx}
\usepackage{amsfonts}
\def\GF{\mathrm {GF}}

\newtheorem{theorem}{Theorem}

\begin{document}
\title{Nearly MDS expander codes with \\ reduced alphabet size}
\author{Marc~A.~Armand,~\IEEEmembership{Senior Member,~IEEE} and Jianwen Zhang}
\maketitle

\begin{abstract}
Recently, Roth and Skachek proposed two methods for constructing nearly
maximum-distance separable (MDS) expander codes. We show that
through the simple modification of using mixed-alphabet codes
derived from MDS codes as constituent codes in their code designs,
one can obtain nearly MDS codes of significantly smaller alphabet size, albeit at the expense of a (very slight) reduction in code rate.
\end{abstract}


\section{Introduction}
In \cite{Guruswami}, Guruswami and Indyk presented a method for
constructing linear-time encodable and decodable nearly
maximum-distance separable (MDS) expander codes. More specifically,
their codes have rate at least $R$ and relative minimum distance
greater than $1-R-\epsilon$ over an alphabet of size $2^{\,{\cal
O}((\log 1/\epsilon)/(R\epsilon^{4}))}$ where $\epsilon>0$ is
sufficiently small and $0<R<1$. Subsequently, \cite{Roth} followed
with two constructions, one yielding linear-time decodable codes,
the other, linear-time encodable and decodable codes, that improved
upon this result by reducing the alphabet size to only $2^{{\cal
O}((1-\epsilon)(\log 1/\epsilon)/\epsilon^{3})}$ and $2^{{\cal
O}(\alpha_{R}(\log 1/\epsilon)/\epsilon^{3})}$, respectively, where $\alpha_{R}$ is dependent on $R$ but may nevertheless be upper bounded by a universal constant independent of $R$.

This correspondence furthers the pursuit of reducing the alphabet size of nearly MDS expander codes. We will show that by using mixed-alphabet codes derived from single-alphabet MDS codes as constituent codes in the constructions by Roth and Skachek, one obtains (i) linear-time decodable nearly MDS codes of rate at least $R-\epsilon$ where $R>\epsilon+1/2$
and with an alphabet size reduced by a factor of about $1-\epsilon$ in the exponent, 
and (ii) linear-time encodable and decodable nearly MDS codes of rate at least $R-\epsilon$ where $R=1-{\cal O}(\epsilon)$
and with an alphabet size reduced by a factor of about $1-{\cal O}(\epsilon)$ in the exponent. We point out that in both cases, the reduction in alphabet size does not come for free and is, in fact, at the expense of a reduction in rate. Nevertheless, we will show that a significant improvement in alphabet size may be achieved at a price of a very small reduction in rate.

\section{Linear-Time Decodable Codes}
\subsection{Preliminaries} \label{Prelim}
Let $G=(V'\!\!:\!\!V'',E)$ be a (finite) $\Delta$-regular, bipartite, undirected, connected graph with edge set $E$ and vertex set $V=V'\cup V''$ such that $|V'|=|V''|=n$ and $V'\cap V''=\{\}$. Each edge in $E$ has one endpoint in $V^{\prime}$ and one in $V^{\prime\prime}$ so that $|E|=n\Delta$. Denote by $E(v)$, the set of $\Delta$ edges incident on vertex $v$, and assume an ordering on $V$ which in turn induces an ordering on the edges in $E(v)$ for each vertex $v\in V$. We associate with each edge $e\in E$, an element of $\GF(2)$, denoted $y_{e}$. In this way, $E$ may be associated with a binary vector of length $n\Delta$ given by $\mathbf{y}=(y_{e})_{e\in E}$ whose entries are indexed by $E$.

Let $(\mathbf{y})_{E(u)}=(y_{e})_{e\in E(u)}$ and denote by $w((\mathbf{y})_{E(u)})$ its Hamming weight. The elements of $\GF(2)$ can easily be assigned to the edges of $G$ such that
\begin{equation} \label{AppendixEqn1}
w((\mathbf{y})_{E(u)})=p\Delta
\end{equation}
for each $u\in V'$ and some fixed $p\in[0,1)$. Trivially, there are many assignment schemes fulfilling this criteria.
However, for a given $\bar{p}$ satisfying $1> \bar{p}\ge p$, it is not immediately clear whether any of these schemes also satisfy
\begin{equation} \label{AppendixEqn2}
w((\mathbf{y})_{E(v)})\leq\bar{p}\Delta
\end{equation}
for each  $v\in V''$. By leveraging on the fact that $G$ is finite, we shall nevertheless prove below that there exist {\em good} assignment schemes, i.e., schemes satisfying both (\ref{AppendixEqn1}) and (\ref{AppendixEqn2}).

\begin{theorem} \label{T1}
For any $p$ and $\bar{p}$ satisfying $0\le p\le \bar{p}< 1$, there exist {\em good} assignment schemes.
\end{theorem}

\begin{proof}
We will show that any assignment scheme which randomly
assigns $p\Delta$ $1$'s to $(\mathbf{y})_{E(u)}$ for each $u\in V'$,
can be adjusted to yield a good assignment scheme by swapping some
$1$'s and $0$'s in $(\mathbf{y})_{E(u)}$ for some $u\in V'$.
We begin by defining the following three types of
vertices in $V''$. Given a vertex $v\in V''$, we say that $v$ is {\em overweight} if $w((\mathbf{y})_{E(v)})>\bar{p}\Delta$, {\em underweight} if $w((\mathbf{y})_{E(v)})<\bar{p}\Delta$, and {\em balanced} otherwise. If there are no overweight vertices, we are done. We therefore focus on the case where there is at least one overweight vertex. Observe that since $\frac{1}{n}\sum_{v\in V''}w((\mathbf{y})_{E(v)})=p\Delta$ and $\bar{p}\geq p$, an overweight vertex implies the existance of at least one underweight vertex.

The graph $G$ may be converted to a directed graph $\overrightarrow{G}=(\overrightarrow{V'}\!\! : \!\!\overrightarrow{V''},\overrightarrow{E})$ as follows. Let $u\in V'$ and $v\in V''$ be two endpoints for an edge $e\in E$. If the bit associated with $e$ is $1$, then the directed edge between $u$ and $v$ is $(v,u)$,
otherwise, it is $(u,v)$. In either case, the directed edge is associated with the same bit as $e$.
In this way, we have a bijective relation between the vertices $\overrightarrow{v}$ (resp., edges $\overrightarrow{e}$) of $\overrightarrow{G}$ and the corresponding vertices $v$ (resp., edges $e$) of $G$.
Moreover, if $u\in V'$, all the out-edges (resp., in-edges) of $\overrightarrow{u}\in\overrightarrow{V'}$ are associated with symbol $0$ (resp., symbol $1$) of $\GF(2)$. On the other hand, if $v\in V''$, all
the out-edges (resp., in-edges) of $\overrightarrow{v}$ are associated with symbol $1$ (resp., symbol $0$). Further, if $v\in V''$ is overweight or underweight or balanced, we say that the corresponding vertex $\overrightarrow{v}$ in $\overrightarrow{G}$ is overweight or
underweight or balanced, respectively.

We next describe a method for reducing the Hamming weight $(\mathbf{y})_{E(v)}$ of an overweight vertex $v\in V''$ by $1$.
\begin{figure}
\centering
\includegraphics[width=3in]{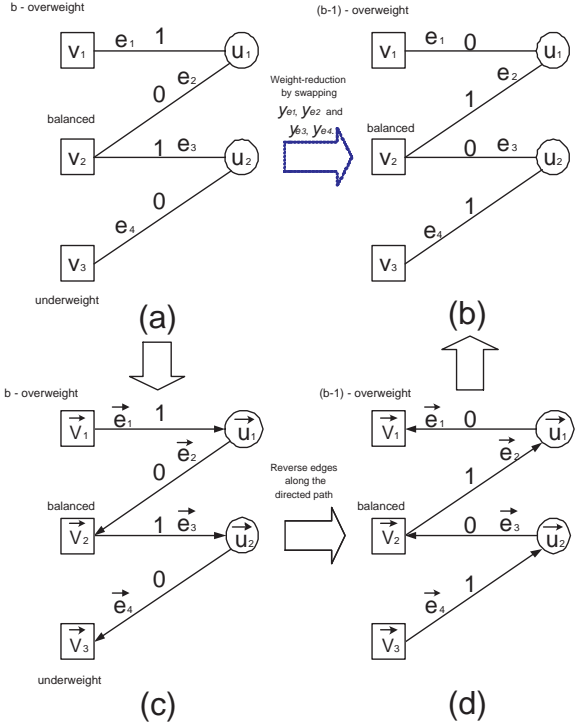}
\caption{Weight-reduction of $(\mathbf{y})_{E(v_{1})}$ owing to overweight vertex $v_{1}\in V''$. (Here, $v_{1}$ is initially $b$-overweight by which we mean that $w((\mathbf{y})_{E(v)})=\bar{p}\Delta+b$.)} \label{fig1}
\end{figure}
Let $u_{1},u_{2}\in V'$ and $v_{1},v_{2},v_{3}\in V''$ as shown in
Fig.\@ \ref{fig1}(a). Now suppose there is
an overweight vertex $\overrightarrow{v_{1}}\in\overrightarrow{V''}$
and a directed path from this vertex to an underweight vertex
$\overrightarrow{v_{3}}\in\overrightarrow{V''}$ (so that $w((\mathbf{y})_{E(v_{3})})<\bar{p}\Delta$), as depicted in Fig.\@ \ref{fig1}(c). Let $w((\mathbf{y})_{E(v_{1})})=\bar{p}\Delta+b$ for some $b\in[1,(1-\bar{p})\Delta]$.
We then ``reverse" all the edges along this directed path as shown in Fig.\@ \ref{fig1}(d) where by reversing an edge, we mean flipping both the direction of a directed edge and the $\GF(2)$ symbol associated with it. The undirected counterpart of Fig.\@ \ref{fig1}(d) is as shown in Fig.\@ \ref{fig1}(b). (Although $\overrightarrow{G}$ changes after an edge reversal, we nevertheless denote the resulting graph by $\overrightarrow{G}$ as well for convenience.)

In obtaining Fig.\@ \ref{fig1}(b), we have effectively swapped $y_{e_{1}}$ and $y_{e_{2}}$ as well as $y_{e_{3}}$ and $y_{e_{4}}$, where $e_{1},e_{2}\in E(u_{1})$ and $e_{3},e_{4}\in E(u_{2})$ such that
the Hamming weight of the resulting $(\mathbf{y})_{E(v_{1})}$ is less than what it was originally by $1$.
Henceforth, we shall refer to this process of reducing $w((\mathbf{y})_{E(v)})$ for $v\in V''$ by $1$ through swapping $1$'s and $0$'s in $(\mathbf{y})_{E(u)}$ for some $u\in V'$ as the {\em weight-reduction} of $(\mathbf{y})_{E(v)}$. In general, an overweight vertex $\overrightarrow{v}\in\overrightarrow{V''}$ with $w((\mathbf{y})_{E(v)})=\bar{p}\Delta+b$ may be converted to a balanced vertex through $b$ weight-reductions of $(\mathbf{y})_{E(v)}$, provided in each instance, the corresponding directed graph $\overrightarrow{G}$ is such that $\overrightarrow{v}$ is linked to an underweight vertex via a directed path. It therefore remains to show that such a directed path exists.

Recall that an overweight vertex in $G$ implies the existance of at least one underweight vertex in the same graph. Correspondingly, if there is an overweight vertex
in $\overrightarrow{G}$, there is at least one vertex underweight in the same graph.
Now suppose some vertex $\overrightarrow{v}\in\overrightarrow{V''}$ is overweight. To
find a directed path linking $\overrightarrow{v}$ to an
underweight vertex, we first follow the out-edges of
$\overrightarrow{v}$. Each of these edges terminates at some vertex
$\overrightarrow{u_{j}}\in \overrightarrow{V'}$. We denote this step by $\overrightarrow{V''}\rightarrow \overrightarrow{V'}$.
Then we return to $\overrightarrow{V''}$ following the out-edges of each $\overrightarrow{u_{j}}$, a step we denote by
$\overrightarrow{V'}\rightarrow \overrightarrow{V''}$. If at least one of these out-edges is incident on an underweight vertex, we are done. If not, the above two steps are repeated until such a vertex is found. Hereafter, we will refer to each pair of consecutive steps, $\overrightarrow{V''}\rightarrow
\overrightarrow{V'}$ and $\overrightarrow{V'}\rightarrow
\overrightarrow{V''}$, as a {\em phase}. We proceed to bound from below, the number of underweight and balanced vertices found at the end of the $(i-1)$th and $i$th phase. This will prove the desired result.

First, assume that no underweight vertex is found in step $\overrightarrow{V'}\rightarrow
\overrightarrow{V''}$ at the end of the $i$th phase. We denote by $S^{(i)}_{b}$ and $S^{(i)}_{o}$, the set of balanced and overweight vertices, respectively, found at the end of this phase, and let $S^{(i)}=S^{(i)}_{o}\cup S^{(i)}_{b}$.
Further, let $T^{(i)}$ be the set of vertices in $\overrightarrow{V'}$ found in step $\overrightarrow{V''}\rightarrow \overrightarrow{V'}$ in the $i$th phase. (The out-edges of the vertices in $T^{(i)}$ are therefore incident on the vertices of $S^{(i)}$.) In step $\overrightarrow{V''}\rightarrow \overrightarrow{V'}$ of phase $i+1$, $|S^{(i)}_{b}|\bar{p}\Delta+\sum_{v_{j}\in S^{(i)}_{o}}p_{j}\Delta$
directed edges are incident on vertices in $\overrightarrow{V'}$, where
\begin{equation}\label{pi}
    p_{j}=\frac{w((\mathbf{y})_{E(v_{j})})}{\Delta}>\bar{p}\ge p.
\end{equation}
Recall that these directed edges are all associated with symbol $1$
and $w((\mathbf{y})_{E(u)})=p\Delta$, where $u\in V'$. Thus, we have
\begin{eqnarray}\label{Ti_low_bound}
\nonumber |T^{(i+1)}| &\ge& \left\lceil\frac{|S^{(i)}_{b}|\bar{p}\Delta+\sum_{v_{j}\in
    S^{(i)}_{o}}p_{j}\Delta}{p\Delta}\right\rceil \\
\nonumber &=&\left\lceil\frac{|S^{(i)}_{b}|\bar{p}+\sum_{v_{j}\in
    S^{(i)}_{o}}p_{j}}{p}\right\rceil \\
   &\ge& |S^{(i)}_{b}|+|S^{(i)}_{o}|+1=|S^{(i)}|+1.
\end{eqnarray}
The last inequality follows from (\ref{pi}). In step $\overrightarrow{V'}\rightarrow
\overrightarrow{V''}$ of phase $i+1$, $|T^{(i+1)}|(1-p)\Delta$ directed edges are incident on vertices in $\overrightarrow{V''}$. If an underweight vertex is reached, we are done. Otherwise, since these directed edges are all
associated with symbol $0$, we have
\begin{eqnarray*}
\! |S^{(i+1)}| \!\!\!\! & \ge & \!\!\!\!\! \left\lceil\frac{|T^{(i+1)}|(1-p)\Delta-\sum_{v_{j}\in S^{(i)}_{o}}(1-p_{j})\Delta}{(1-\bar{p})\Delta}+|S^{(i)}_{o}| \! \right\rceil \\
\!\!\!\!\! & = & \!\!\!\!\! \left\lceil|T^{(i+1)}|\frac{1-p}{1-\bar{p}}+\frac{\sum_{v_{j}\in
  S^{(i)}_{o}}(p_{j}-\bar{p})}{1-\bar{p}}\right\rceil.
\end{eqnarray*}
This lower bound may be achieved when no overweight vertices, in addition to those already contained in $S_{o}^{(i)}$, are found in step $\overrightarrow{V'}\rightarrow\overrightarrow{V''}$ of phase $i+1$. The right-hand-side may be further lower-bounded by
$$|T^{(i+1)}|+1$$
since $1-\bar{p}\le 1-p$ and $p_{j}>\bar{p}$. From (\ref{Ti_low_bound}), we have
$$|S^{(i+1)}|\ge |S^{(i)}|+2.$$
Hence, provided no underweight vertex is found, $|S^{(i)}|$ increases with $i$. Since $\overrightarrow{G}$ is finite, it follows that a directed path linking an overweight vertex to an underweight one will eventually be found after some finite number of phases, as desired.

That there may be more than one good assignment scheme is obvious.
\end{proof}

Now consider a related problem. Let $F_{1}$ and $F_{2}$ be two alphabets.
For a given choice of $p$, it is desired that for each vertex $u\in V'$, exactly $p\Delta$ edges in $E(u)$ be associated with $F_{1}$ and the remaining edges in $E(u)$ be associated with $F_{2}$. It is further desired that for each vertex $v\in V''$, at most $R\Delta$ edges in $E(v)$ be associated with $F_{1}$ and the remaining edges in $E(v)$ be associated with $F_{2}$ for some fixed $R\in(p,1]$.

Trivially, when $R=1$, the above two conditions are easily satisfied. When $R<1$ however, the $p\Delta$ edges of $E(u)$ to be associated with $F_{1}$ have to be appropriately chosen for each $u\in V'$ so that the second of the two conditions above may be satisfied for each $v\in V''$. By Theorem \ref{T1}, this can always be achieved. Henceforth, we assume knowledge of a good assignment scheme given by a vector $\mathbf{y}^{\ast}\in\GF(2)^{n\Delta}$. For each vertex $u\in V'$, the edges in $E(u)$ to be associated with $F_{1}$ are then given by the support of $(\mathbf{y}^{\ast})_{E(u)}$.

\subsection{The Modified Construction}
For the remainder of this paper, let $F_{1}=\GF(q_{1})$ be a proper subfield of $F_{2}=\GF(q_{2})$ and $\alpha$ be a rational  number satisfying $0<\alpha<1$ and $q_{1}=q_{2}^{\alpha}$. For convenience, we denote by $F_{1}$ and $F_{2}$, the respective additive groups of $\GF(q_{1})$ and $\GF(q_{2})$ as well.

Let ${\cal C}^{\prime}$ and ${\cal C}^{\prime\prime}$ be
$(\Delta,r\Delta,\theta\Delta)$ and $(\Delta,R\Delta,\delta\Delta)$
MDS codes over $F_{2}$, respectively. For some fixed $p\leq r$ and
$u\in V'$, let $\Phi_{\,u}$ be a direct product of groups involving
$p\Delta$ copies of $F_{1}$ and $(r-p)\Delta$ copies of $F_{2}$. Next, let
$\varepsilon_{u}:\Phi_{u}\rightarrow{\cal C}_{u}^{\prime}$ be a
systematic encoding function for the subcode ${\cal
C}_{\,u}^{\;\prime}$ of ${\cal C}^{\prime}$ obtained by encoding the
subset $\Phi_{u}$ of information words from $F_{\,2}^{\;r\Delta}$. The information-bearing code coordinates defined over $F_{1}$ are given by the support of $(\mathbf{y}^{\ast})_{E(u)}$ where $u\in V'$.
(Thus, the information-bearing coordinates are not necessarily among the first or last $r\Delta$ code coordinates.)
From \cite[Theorem 5]{Sidorenko}, the relative minimum distance of
${\cal C}_{u}^{\,\prime}$ is also $\theta$. Further, its rate
$r^{\,\prime}$ is
$$\frac{\log_{q_{2}}\left|{\cal C}_{u}^{\prime}\right|}{\log_{q_{2}}
\left|F_{1}^{p\Delta}F_{2}^{(1-p)\Delta}\right|}=\frac{p(\alpha-1)+r}{p(\alpha-1)+1}$$
and so is independent of $u$. Similarly, let ${\cal
C}_{v}^{\,\prime\prime}$ be the subcode of ${\cal C}^{\prime\prime}$
of relative minimum distance $\delta$ and rate
$R_{v}^{\,\prime}=(j_{v}(\alpha-1)+R)/(j_{v}(\alpha-1)+1)$ obtained
by systematically encoding the subset
$F_{1}^{\,j_{v}\Delta}F_{2}^{(R-j_{v})\Delta}$ of information words
from $F_{2}^{\,R\Delta}$ for some $j_{v}\leq R$. Moreover, the information-bearing code coordinates defined over $F_{1}$ are given by the support of $(\mathbf{y}^{\ast})_{E(v)}$ where $v\in V''$.

Now, let $\mathbf{x}=(x_{e})_{e\in E}$ be a word of length $|E|$ whose
entries are indexed by $E$ with $p\Delta n$ (resp., $(1-p)\Delta n$)
of them being elements of $F_{\,1}$, (resp., $F_{2}$). Further, let
$(\mathbf{x})_{E(v)}$ be the subblock of $\mathbf{x}$ indexed by
$E(v)$. As in \cite{Barg}, we define the code $C$ to be the
collection of such mixed-alphabet words such that for each codeword
$\mathbf{c}\in C$, $(\mathbf{c})_{\,E(u)}$ (resp.,
$(\mathbf{c})_{E(v)}$) is a codeword of ${\cal
C}_{h_{u}}^{\,\prime}$ (resp., ${\cal C}_{h_{v}}^{\prime\prime}$)
for every $u\in V^{\,\prime}$ (resp., $v\in V^{\prime\prime}$).
Next, following \cite{Roth}, let
$\Phi=\prod_{u\in V'}\Phi_{u}$ and $\psi:
C\rightarrow\Phi$ be given by
$\psi(\mathbf{c})=\left(\varepsilon_{u}^{-1}\left((\mathbf{c})_{E(u)}\right)\right)_{u\in
V^{\prime}}$ where $\mathbf{c}\in C$, and define the code
$(C)_{\Phi}\subset\Phi$ by $(C)_{\Phi}=\{\psi(\mathbf{c}): \;
\mathbf{c}\in C\}$. Thus, the code $C$ may be viewed as a concatenated
code with $(C)_{\,\Phi}$ as the outer code and $n$ inner codes,
i.e., the ${\cal C}_{\,u}^{\,\prime}$ for $u\in V^{\prime}$,
each of rate $r^{\,\prime}$.

Trivially, when $p=0$, the alphabet size $|\Phi_{u}|$ of $(C)_{\,\Phi}$ is simply $\phi=q_{1}^{p\Delta}q_{2}^{(r-p)\Delta}=q_{2}^{r\Delta}$, coinciding with the alphabet size of the code $(C)_{\,\Phi}$ in \cite[Section II]{Roth}. More generally, $|\Phi_{u}|=\phi^{1-(p(1-\alpha)/r)}$.
We proceed to obtain a lower bound on the rate of $(C)_{\,\Phi}$.

\subsection{Bounds on the Parameters of $(C)_{\Phi}$}
The length $N_{{\cal C}_{u}^{\prime}}$ (resp., $N_{{\cal
C}_{v}^{\prime\prime}}$) of ${\cal C}_{u}^{\prime}$ (resp., ${\cal
C}_{v}^{\prime\prime}$) in $q_{2}$-ary symbols is
$(p(\alpha-1)+1)\Delta$, (resp., $(j_{v}(\alpha-1)+1)\Delta$). The
redundancy of ${\cal C}_{u}^{\prime}$ (resp., ${\cal
C}_{v}^{\prime\prime}$) in $q_{2}$-ary symbols is therefore
$N_{\,{\cal C}_{u}^{\prime}}(1-r^{\prime})$, (resp., $N_{{\cal
C}_{v}^{\prime\prime}}(1-R_{v}^{\prime})$). Consequently, if
$R_{\,C}$ denotes the rate of $C$, then its redundancy in
$q_{2}$-ary symbols is $nN_{{\cal C}_{u}^{\prime}}(1-R_{C})$.
Following \cite{Barg}, since $C$ is additive (so that any linear
combination of codewords is also a codeword) and its redundancy is
at most the sum of the redundancies of its constituent codes, we
have that
$$nN_{{\cal C}_{u}^{\prime}}(1-R_{C})\leq nN_{{\cal C}_{u}^{\prime}}
(1-r^{\prime})+\sum_{v\in V''}N_{{\cal
C}_{v}^{\prime\prime}}(1-R_{v}^{\prime})$$ and after some
substitutions and rearrangement, obtain
$$R_{C}\geq r^{\prime}+\frac{p(\alpha-1)+R}{p(\alpha-1)+1}-1.$$
Recalling the concatenated code structure of $C$ described above, it
follows that the rate of $(C)_{\Phi}$ is bounded from below by
\begin{equation} \label{Lowerbound_Rate}
1-\frac{1}{r^{\prime}}+\frac{p(\alpha-1)+R}{p(\alpha-1)+r}=1+\frac{R-1}{p(\alpha-1)+r}.
\end{equation}

We next turn to the relative minimum distance of $(C)_{\,\Phi}$.
First, observe that the proofs of \cite[Lemma 3.2 \& Proposition
3.3]{Roth} are independent of the underlying code alphabet.
Moreover, recall that for each $u\in V'$ and $v\in V''$, the constituent codes ${\cal C}_{\,u}^{\,\prime}$ and
${\cal C}_{v}^{\prime\prime}$ have the same relative minimum
distance as their respective ``parent" codes, i.e., ${\cal
C}^{\,\prime}$ and ${\cal C}^{\prime\prime}$. Consequently,
\cite[Theorem 3.1]{Roth} remains applicable in the current setting,
with the proof requiring only minor changes, mainly notational. From
this theorem, we have that the relative minimum distance of
$(C)_{\Phi}$ is bounded from below by
\begin{equation} \label{Bound_distance}
\frac{\delta-\gamma_{G}\sqrt{\delta/\theta}}{1-\gamma_{G}}
\end{equation}
where $\gamma_{G}$ is the ratio of the second largest eigenvalue of
the adjacency matrix of $G$ to its largest, i.e., $\Delta$.

\subsection{Obtaining Nearly MDS Codes} \label{ObtainingNearlyMDSCodes}
Following \cite[Example 3.1]{Roth}, let $\theta=\epsilon$ for some
small $\epsilon\in (0,1]$ so that $r>1-\epsilon$ and let
$q_{2}>\Delta\geq 4/\epsilon^{3}$ and $G$ be Ramanujan. Following
that example, the relative minimum distance of $(C)_{\,\Phi}$ is at
least
$$\frac{\delta-\gamma_{G}\sqrt{\delta/\theta}}{1-\gamma_{G}}>1-R-\epsilon.$$
Next, let
\begin{equation} \label{Lowerbound_R}
R>\epsilon+1/2
\end{equation}
and for a given choice of $\alpha$,
\begin{equation} \label{Upperbound_p}
p\leq\epsilon/(1-\alpha)
\end{equation}
By (\ref{Lowerbound_Rate}), the rate of $(C)_{\Phi}$ is bounded from
below by
\begin{eqnarray*}
1+\frac{R-1}{p(\alpha-1)+1-\epsilon} & = & \frac{p(\alpha-1)+R-\epsilon}{p(\alpha-1)+1-\epsilon} \\
& \geq & \frac{R-2\epsilon}{1-2\epsilon} \, , \;\; \mbox{ by (\ref{Upperbound_p})} \\
& > & R-\epsilon \, , \;\; \mbox{ by (\ref{Lowerbound_R})}.
\end{eqnarray*}
Thus, $(C)_{\Phi}$ approaches the Singleton bound as
$\epsilon\rightarrow 0$. Nevertheless, due to (\ref{Lowerbound_R}),
note that the above code design yields only codes of moderate to
high rate. In particular, if $p(1-\alpha)=\epsilon$, then
\begin{eqnarray*}
|\Phi_{u}| & = & \phi^{1-\frac{\epsilon}{r}} \\
& \leq & \phi^{1-\frac{\epsilon}{1+(\epsilon^{3}/4)-\epsilon}} \, , \;\;  \mbox{ since } 1/\Delta\leq\epsilon^{3}/4 \\
& \leq & \phi^{1-\epsilon}
\end{eqnarray*}
where the bound $1-\frac{\epsilon}{1+(\epsilon^{3}/4)-\epsilon}\leq 1-\epsilon$ is tight since $\epsilon$ is small.

Finally, as noted in \cite{Sidorenko}, the codes ${\cal
C}_{u}^{\prime}$ and ${\cal C}_{v}^{\prime\prime}$ may be decoded
using the decoders for their respective ``parent" codes. With that,
it is clear that $(C)_{\Phi}$ may be decoded using the linear-time
decoder in \cite[Fig.\@ 1]{Roth}.

\subsection{Comparisons to the Original Construction} \label{SecII-E}
It should be noted that the rate of the nearly MDS code in \cite[Example 3.1]{Roth} resulting from the original construction of \cite[Section II]{Roth}, is bounded from below by
$$1-\frac{1}{1-\epsilon}+\frac{R}{1-\epsilon}=\frac{R-\epsilon}{1-\epsilon}>R-\epsilon/2$$
when $R$ satisfies (\ref{Lowerbound_R}). On the other hand, our modifications to this construction yields a nearly-MDS code of rate exceeding $R-\epsilon$.

Thus, the reduction in alphabet size from $\phi$ to $\phi^{1-(\epsilon/r)}$ (which recall, is achieved when $p(1-\alpha)=\epsilon$), is at the expense of a reduction in rate. In other words, what we have is {\em not} an improvement of the results of \cite{Roth} but rather, a trade-off between alphabet size and rate.

Nevertheless, we have that for $R$ satisfying (\ref{Lowerbound_R}),
$$\frac{(R-\frac{\epsilon}{2})-(R-\epsilon)}{R-\frac{\epsilon}{2}}=\frac{\epsilon}{2R-\epsilon}<\frac{\epsilon}{2(\frac{1}{2}+\epsilon)-\epsilon}=\frac{\epsilon}{1+\epsilon}<\epsilon$$
and for $p(1-\alpha)=\epsilon$, $\Delta=4/\epsilon^{3}$ and $q_{2}={\cal O}(1/\epsilon^{3})$,
$$\frac{\phi-\phi^{1-(\epsilon/r)}}{\phi} 
=1-q_{2}^{-\Delta\epsilon}=1-{\cal O}(\epsilon^{3})^{4/\epsilon^{2}}\rightarrow 1$$
as $\epsilon\rightarrow 0$. Thus, the trade-off between alphabet size and rate is remarkable, for a significant reduction in alphabet size is achievable at a price of a very small reduction in rate.

\section{Linear-Time Encodable \& Decodable Codes}
We now turn to the linear-time encodable code construction method presented in \cite[Section V]{Roth}. In this construction, two bipartite regular graphs $G_{1}$ and $G_{2}$ as well as four constituent codes are needed, three of which being MDS codes over $F_{2}$ with parameters $[\Delta_{1}, r_{0}\Delta_{1}]$, $[\Delta_{1}, R_{1}\Delta_{1}]$ and $[\Delta_{2}, R_{2}\Delta_{1}]$. The forth code is a $(n,r_{m}n)$ code over $F_{2}^{R_{2}\Delta_{2}}$ which, as stated in \cite{Roth}, could be the code of \cite{Spielman}. We proceed to describe a simple modification involving only the graph $G_{1}$ and the $[\Delta_{1}, R_{1}\Delta_{1}]$ constituent code which will ultimately lead to linear-time encodable and decodable nearly MDS codes over smaller alphabets.

\subsection{The Proposed Modification}
Let the graph $G_{1}=(V':V'',E_{1})$ have degree $\Delta_{1}$  such that $|V'|=|V''|=n$ as before. Denote the set of
edges incident on a vertex $u$ in $G_{1}$ by $E_{1}(u)$. As in Section \ref{Prelim}, assume an ordering on the vertex set $V'\cup V''$ of $G_{1}$ which in turn induces an ordering on the edges in $E_{1}(v)$ for each vertex $v\in V''$.
Moreover, denote the $[\Delta_{1}, R_{1}\Delta_{1}, \delta_{1}\Delta_{1}]$ constituent code by $\mathcal{C}_{1}$.

For some fixed $p\in [0,R_{1}]$, we associate appropriate $p\Delta_{1}$ edges in
$E_{1}(v)$ with $F_{1}$ and the remaining edges in $E_{1}(v)$ with
$F_{2}$ for each $v\in V''$ such that exactly $p\Delta_{1}$ edges in
$E_{1}(u)$ are associated with $F_{1}$ for each $u\in V'$.
A good assignment scheme, denoted $\mathbf{y^{\ast}}\in \GF(2)^{n\Delta_{1}}$, is therefore needed. By Theorem \ref{T1}, such a scheme can always be found. For each $v\in V''$, the edges in $E_{1}(v)$ to be associated with $F_{1}$ are then given by the
support of $(\mathbf{y}^{\ast})_{E_{1}(v)}$.

For each $v\in V''$, let $\Omega_{v}$ be a direct product of groups involving $p\Delta_{1}$ copies of $F_{1}$ and $(R_{1}-p)\Delta_{1}$ copies of $F_{2}$. In addition, let $\varepsilon_{1,v}: \Omega_{v}\rightarrow\mathcal{C}_{1,v}$ be a systematic encoding function  for the subcode $\mathcal{C}_{1,v}$ of $\mathcal{C}_{1}$ obtained by encoding the subset $\Omega_{v}$ of information words in $F_{2}^{R_{1}\Delta_{1}}$. The information-bearing code coordinates of $\mathcal{C}_{1,v}$ defined over $F_{1}$ are given by the support of $(\mathbf{y}^{\ast})_{E_{1}(v)}$. From \cite[Theorem 5]{Sidorenko}, the relative minimum distance of ${\cal C}_{1,v}$ is also $\delta_{1}$ for each $v\in V''$.

Observe that if $\mathbf{c}$ is an $n\Delta_{1}$-tuple with $p\Delta_{1}$ elements defined over $F_{1}$ and $(1-p)\Delta_{1}$ elements defined over $F_{2}$ such that $(\mathbf{c})_{E_{1}(v)}\in{\cal C}_{1}$ for each $v\in V''$, then for each $u\in V'$, exactly $p\Delta_{1}$ (resp.\@ $(1-p)\Delta_{1}$) elements of the $\Delta_{1}$-tuple $(\mathbf{c})_{E_{1}(u)}$ are over $F_{1}$ (resp.\@ $F_{2}$). With that, the construction in \cite[Section V]{Roth} may be modified to obtain a linear-time encodable code $\mathbb{C}$ of smaller alphabet size by replacing the encoding function $\varepsilon_{1}$ in Step E1 of \cite[Fig.\@ 2]{Roth} by the $\varepsilon_{1,v}$, while keeping Steps E2 to E4 unchanged. More specifically, given an information word from $\prod_{v\in V''}\Omega_{v}$, the resulting encoder outputs the codeword $\mathbf{x}=(\mathbf{x}_{u})_{u\in V'}$ where each $\mathbf{x}_{u}$ is an element of an alphabet of size
$$\Gamma=q_{1}^{p\Delta_{1}}q_{2}^{(1-p)\Delta_{1}+\Delta_{2}}.$$

\subsection{Obtaining Nearly MDS Codes}
Following \cite[Section V-C]{Roth}, let
$(1-r_{0})\Delta_{1}=r_{m}R_{2}\Delta_{2}$ and $R=R_{1}=R_{2}$. Further, let $\kappa$ be a universal constant such that $r_{m}\ge \kappa$  and $1-r_{0}<\kappa\epsilon$. For this choice of values,
$$\Gamma=q_{2}^{(1-s+\frac{1-r_{0}}{r_{m}R})\Delta_{1}}$$
where $s=p(1-\alpha)$. On the other hand, the corresponding code constructed in \cite[Section V-C]{Roth} has alphabet size
$$\gamma=q_{2}^{\Delta_{1}+\Delta_{2}}=q_{2}^{(1+\frac{1-r_{0}}{r_{m}R})\Delta_{1}}.$$
Now
$$\frac{1-s+\frac{1-r_{0}}{r_{m}R}}{1+\frac{1-r_{0}}{r_{m}R}}=1-\frac{s}{1+\frac{1-r_{0}}{r_{m}R}}
<1-\frac{R}{R+\epsilon}s$$
since $\frac{1-r_{0}}{r_{m}R}<\frac{\epsilon}{R}$. Thus, $\Gamma<\gamma^{1-\frac{R}{R+\epsilon}s}$ and so a larger value of $s$ is clearly desirable. We will show that when $R=1-{\cal O}(\epsilon)$ or $R={\cal O}(\epsilon)$, $\mathbb{C}$ is nearly MDS with alphabet size less than $\gamma^{1-{\cal O}(\epsilon)}$ and rate greater than $R-\epsilon$.

We begin with the rate $r_{\mathbb{C}}$ of $\mathbb{C}$ which may be expressed as
$$\frac{\log_{q_{2}}|\Omega|^{n}}{\log_{q_{2}}\Gamma^{n}}=\frac{(R-s)\Delta_{1}n}{(1-s)\Delta_{1}n+\Delta_{2}n}=\frac{(R-s)\Delta_{1}}{(1-s)\Delta_{1}+\Delta_{2}}.$$
Since
$\Delta_{2}=\frac{(1-r_{0})}{r_{m}R}\Delta_{1}<\frac{\epsilon}{R}\Delta_{1}$, it follows that
\begin{equation}\label{Sect3_coderate}
r_{\mathbb{C}}>\frac{R-s}{1-s+\frac{\epsilon}{R}}.
\end{equation}
We wish to have
$\frac{R-s}{1-s+\frac{\epsilon}{R}}\ge R-\epsilon$, or equivalently
\begin{equation}\label{Range_r}
    sR^2-s(1+\epsilon)R+\epsilon^2\ge 0
\end{equation}
from which, we obtain the bound
\begin{equation}\label{s_upperbound}
    s\le \frac{\epsilon^2}{R(1+\epsilon-R)}.
\end{equation}
If the discriminant $s^2(1+\epsilon)^2-4s\epsilon^2$ of the left-hand-side of (\ref{Range_r}) is less than zero, then
$$0<s<\frac{4\epsilon^2}{(1+\epsilon)^2}$$
otherwise
\begin{equation} \label{limits1}
\frac{4\epsilon^2}{(1+\epsilon)^2}\le s\le\frac{\epsilon^2}{R(1+\epsilon-R)}
\end{equation}
where the last inequality follows from (\ref{s_upperbound}) and the fact that $\frac{4\epsilon^2}{(1+\epsilon)^2}\le\frac{\epsilon^2}{R(1+\epsilon-R)}$. Since a larger value of $s$ is desirable, we focus on the latter case for which, $r_{\mathbb{C}}>R-\epsilon$ when
\begin{equation}\label{R_highrate}
    R\ge\frac{1+\epsilon+\sqrt{(1+\epsilon)^2-\frac{4\epsilon^2}{s}}}{2}
\end{equation}
or
$$R\le\frac{1+\epsilon-\sqrt{(1+\epsilon)^2-\frac{4\epsilon^2}{s}}}{2}.$$

Therefore, for $R\in(0,1]$, we have that
$$1-\frac{\epsilon^2}{(R+\epsilon)(1+\epsilon-R)}\le 1-\frac{R}{R+\epsilon}s\le
1-\frac{4R\epsilon^2}{(R+\epsilon)(1+\epsilon)^{2}}.$$
In particular, when $R=1-{\cal O}(\epsilon)$ or $R={\cal O}(\epsilon)$,
we have that
$$\Gamma<\gamma^{1-{\cal O}(\epsilon)}.$$

Finally, we consider the minimum distance of $\mathbb{C}$. Since
$\mathcal{C}_{1}$ can be decoded by the decoder of its ``parent"
code, the decoding algorithm in \cite[Fig.\@ 4]{Roth} can be applied
to $\mathbb{C}$. Provided both $G_{1}$ and $G_{2}$ are Ramanujan, any received word with $t$ errors and $\rho$
erasures such that $2t+\rho\le (1-R-\epsilon)n$ can be uniquely decoded to the transmitted codeword. This implies that the relative minimum distance of $\mathbb{C}$ is at least $1-R-\epsilon$ and so $\mathbb{C}$ is nearly MDS.

\subsection{Comparisons to the Original Construction}
As with the linear-time decodable code construction, the reduction
in alphabet size from $\gamma$ to $\Gamma$ is at the expense of a
reduction in rate, as the rate of $\mathbb{C}$ and its counterpart
in \cite[Section V]{Roth} is lower bounded by
$\frac{R-s}{1-s+\frac{\epsilon}{R}}$ (from (\ref{Sect3_coderate}))
and $\frac{R}{1+\frac{\epsilon}{R}}$, respectively, and
$$\frac{R}{1+\frac{\epsilon}{R}}-\frac{R-s}{1-s+\frac{\epsilon}{R}}=\frac{s(1-R)+\frac{s\epsilon}{R}}{(1+\frac{\epsilon}{R})(1-s+\frac{\epsilon}{R})}>0.$$
Nevertheless, as in Section \ref{SecII-E}, we will show that the reduction in rate our modification brings, is marginal compared to the improvement in alphabet size in the high rate case at least, i.e., when $R$ satisfies (\ref{R_highrate}).

First, since $sR>0$, we have that
$$\frac{\frac{R}{1+\frac{\epsilon}{R}}-\frac{R-s}{1-s+\frac{\epsilon}{R}}}{\frac{R}{1+\frac{\epsilon}{R}}}=\frac{s+\frac{\epsilon}{R}-sR}{R+\epsilon-sR}<\frac{s+\frac{s\epsilon}{R}}{R+\epsilon}=\frac{s}{R}.$$
When $R$ satisfies (\ref{R_highrate}), it follows from (\ref{limits1}) that $s$ and in turn $\frac{s}{R}$, tend to zero as $\epsilon\rightarrow 0$. On the other hand, setting $\Delta_{1}=\alpha_{R}/\epsilon^3$ as in~\cite[Section V-C]{Roth}, we have that
$$s\Delta_{1}\ge \frac{4\epsilon^2}{(1+\epsilon)^2}\frac{\alpha_{R}}{\epsilon^3}=\frac{4\alpha_{R}}{\epsilon(1+\epsilon)^2}$$
and so when $q_{2}={\cal O}(1/\epsilon^3)$,
$$\frac{\gamma-\Gamma}{\gamma}=1-q_{2}^{-s\Delta_{1}}\ge 1-{\cal O}(\epsilon^3)^{\frac{4\alpha_{R}}{\epsilon(1+\epsilon)^2}}\rightarrow 1$$
as $\epsilon \rightarrow 0$. Finally, since $(\gamma-\Gamma)/\gamma<1$, it follows that $(\gamma-\Gamma)/\gamma\rightarrow 1$ as $\epsilon\rightarrow 0$ and so once again, we see that a significant reduction in alphabet size is attainable at a price of a very small reduction in rate.

\end{document}